\documentclass[reprint,aps,pra,twocolumn,showpacs,showkeys,superscriptaddress,
amssymb]{revtex4}

\usepackage[pdftex]{graphicx}
\usepackage{hyperref}

\usepackage{amssymb}
\newcommand{\ep}{\varepsilon}
\newcommand{\dl}{\delta}
\newcommand{\Tr}{\textrm{Tr}}

\newcommand{\rmi}{\textrm{i}}
\newcommand{\rmf}{{\rm f}}
\newcommand{\rme}{{\rm e}}
\newcommand{\su}{{\mathfrak{su}}}

\begin{document}
\title{Trap-free manipulation in the Landau-Zener system}
\author{Alexander Pechen}
\email{apechen@gmail.com}
\homepage{http://www.mathnet.ru/eng/person17991}
\affiliation{Department of Chemical Physics, Weizmann Institute of Science,
Rehovot 76100, Israel}
\affiliation{Steklov Mathematical Institute of Russian
Academy of Sciences, Gubkina 8, Moscow 119991, Russia}
\author{Nikolay Il'in}
\affiliation{Steklov Mathematical Institute of Russian
Academy of Sciences, Gubkina 8, Moscow 119991, Russia}
\date{\today}

\begin{abstract}
The analysis of traps, i.e., locally but not globally optimal controls, for
quantum control systems has attracted a great interest in recent years. The
central problem that has been remained open is to demonstrate for a given system
either existence or absence of traps. We prove the absence of traps and hence
completely solve this problem for the important tasks of unconstrained
manipulation of the transition probability and unitary gate generation in the
Landau-Zener system---a system with a wide range of applications across physics,
chemistry and biochemistry. This finding provides the first example of a
controlled quantum system which is completely free of traps. We also discuss the
impact of laboratory constraints due to decoherence, noise in
the control pulse, and restrictions on the available controls which when being
sufficiently severe can produce traps.
\end{abstract}
\maketitle

\section{Introduction}
Manipulation by atomic and molecular systems is an important branch of modern
science with applications ranging from optimal laser driven population transfer
in atomic systems out to laser assisted control of chemical reactions
\cite{Tannor1985}. High interest is directed towards control of the
Landau-Zener (LZ) system---a two-state quantum system whose unitary evolution
under the action of
the
control $\ep(t)$ (e.g., shaped laser field) is governed by the equation
\begin{equation}\label{eq1:2}
\dot U^\ep_t=-\rmi(\Delta\sigma_x+\ep(t)\sigma_z)U^\ep_t,\qquad
U^\ep_{t=0}=\mathbb I
\end{equation}
where $\Delta>0$, $\sigma_x$ and $\sigma_z$ are the Pauli matrices. The case
$\ep(t)=\ep t$ with constant $\ep$ was studied by Landau, Zener,
St\"uckelberg,
and Majorana~\cite{Landau1932}. This system has been widely applied in physics,
chemistry, and biochemistry, e.g., for describing transfer of charge along with
its energy~\cite{Kuztetsov}, photosynthesis~\cite{Warshel1980}, atomic and
molecular collisions, processes in plasma physics~\cite{Smirnov}, Bose-Einstein
condensate~\cite{Bason2012}, experimental
realizations of qubits,
etc.~\cite{Pokrovsky2009,Shore1992,Arimondo2003,Caneva2009,Brataas2011,
Baturina2011,Avron2011}.

Controlled manipulation by a quantum system can be formulated as finding global
maxima of a suitable objective $J(\ep)$ associated to the system. For example,
maximizing the probability of transition from the initial state $|\rmi\rangle$
to a target final state $|\rmf\rangle$ at a final time $T$ can be described by
maximizing $J(\ep)=P_{\rmi\to\rmf}=|\langle\rmf|U^\ep_T|\rmi\rangle|^2$. A
control which attains a local maximum of $J$ can be found either
numerically using the model of the system or experimentally. In
both circumstances, the first step of a common procedure is
to apply a trial pulse $\ep_0$ and obtain the outcome $J(\ep_0)$, either
numerically or measuring it in the laboratory. The second step is to make
various small modifications of $\ep_0$ and find $\ep_1$ which
produces maximum increase in $J$. Then $\ep_1$ is used as a new trial pulse and
the procedure is repeated until no
significant increase is produced or a maximum number of iterations is reached.

Of crucial practical importance is to know whether $J(\ep)$ has {\it traps},
i.e. local maxima with the values less than the global maximum, as necessary to
properly choose between local (e.g., gradient) and global optimization
methods~\cite{Tannor1992,Krotov2009,Fradkov2005,GRAPE,Eitan2011,Machnes2011}.
Traps can
strongly influence on both theoretical and experimental quantum control
studies---they determine the level of difficulty of controlling the system and
can significantly slow down or even completely prevent finding globally optimal
controls. Whereas the analysis of traps in manipulation by quantum systems has
attracted high attention~\cite{NOTRAPS1,Ho2006,NOTRAPS5,Ruetzel2010,Wu2009,
Schirmer2010,PechenTannor2011,NumericalResults,PechenTannor2012}, no examples of
trap-free quantum systems have been known. Only partial theoretical results have
been obtained stating the absence of traps at special regular controls. This
finding does not at all exclude the absence of traps that makes the problem open
since even a single trap may produce significant difficulties for the
optimization if it has a large attracting domain~\cite{PechenTannorReply}. In
this work we show that the LZ system is trap-free and hence, for
example, the only extrema of $J(\ep)=P_{\rmi\to\rmf}$ for this system are global
maxima and minima. This finding provides the first example of a trap-free
quantum control system where unconstrained local manipulations are always
sufficient to find best control pulses. Sufficiently strong constraints on the
controls may destroy this property and in the end we discuss possible
limitations for the analysis due to decoherence, noise in the control
pulses and limited tunability of the control strength and time scales in
laboratory experiments.

\section{Traps and control landscapes}
Formally, a control field $\ep(t)$ is a
trap for the objective $J(\ep)$ if it is a local maximum, i.e., a
maximum with the value less than the global maximum, $J(\ep)<J_{\rm
max}=\max\limits_\ep J(\ep)$ (in this work we consider as control goal
maximizing the objective; if the goal is to minimize the objective then
traps are local minima). Answering the question whether traps exist for a
given control problem is crucial for determining proper algorithms and our
abilities for finding optimal control fields. In the absence of traps, local
search algorithms should generally be able to find globally optimal controls
(exceptions may occur if the initial control is chosen exactly at a saddle
point, where the gradient of the objective is zero). If the objective has traps
(perhaps even a single trap with large attracting domain) then local search
procedures may converge to local maxima instead of attaining a desired globally
optimal control and more sophisticated global search methods should be
exploited for a successful optimization.

Traps are critical points, i.e., the gradient $\nabla J_\ep=0$ at any trap.
Critical points for control objectives $J(\ep)=P_{\rmi\to\rmf}$ were studied in
seminal works~\cite{NOTRAPS1}, where the absence of traps was suggested. The
suggestion was drawn from the proof that any function of the form
$f(U)=\Tr[U\rho_0U^\dagger O]$ ($\rho_0$ is a positive matrix and $O$ is
Hermitian) defined on the unitary group $U(n)$, where $n$ is the system
dimension, has as extrema only global maxima, global minima, and saddles and has
no traps. (Extrema of trace functions over unitary and orthogonal groups
were studied in other contexts by J. von Neumann~\cite{Neumann},
R.~Brockett~\cite{Brockett}, S.~Glaser \emph{et al}~\cite{Glaser}, etc.) Then,
under the controllability condition which assumes that any $U\in U(n)$ can be
generated by some control, this result was used to conclude the absence of traps
for the underlying objective functional $J(\ep)$. Later it was shown that the
conclusion of the absence of traps requires an additional assumption that the
map $\chi:\ep\to U^\ep_T$ is non-degenerate~\cite{Ho2006}, meaning that
arbitrary
infinitesimal variations of $\ep$ produce variations of $U^\ep_T$ in all
directions on $U(n)$~\cite{Raj2007,Wu2008}. While the controllability condition
is relatively easy to verify~\cite{Controllability1}, checking the
non-degeneracy assumption turned out to be a hard problem. Moreover, critical
controls
violating this assumption were found~\cite{Wu2009,Schirmer2010},
and even
second-order traps---critical controls which are not global maxima and where the
Hessian $H=\dl^2 J/(\dl\ep)^2$ is negative semidefinite were shown to exist
under rather general assumptions~\cite{PechenTannor2011}. (Second-order traps
are not necessarily local maxima but effectively they are traps for local
algorithms exploiting at most second order local information about the
objective; see Chapter~20 of~\cite{Agrachev2004} for a general discussion of the
non-degeneracy and second order optimality conditions.) These findings led to
reconsideration of the conclusion of absence of traps. Some numerical
simulations suggested that the condition of non-degeneracy might be generally
satisfied or at least its violation does not produce multiple
traps~\cite{NumericalResults}, while other indicated possible trapping
behavior~\cite{Schirmer2010,PechenTannorReply}. However, numerical search is
limited and the extent to which these runs span the full space of quantum
control possibilities is questionable~\cite{PechenTannorReply}. Hence the
problem of proving either existence or absence of traps has been remained open.

\section{Absence of traps for the Landau-Zener system}
Our main result is that
the only critical points of any objective of the form
$J(\ep)=f(U^\ep_T)$, where $U^\ep_T$
satisfies~(\ref{eq1:2}) and $f(U)$ is any function on the special unitary
group $SU(2)$ which has no local extrema, are global maxima, global minima,
and the zero control field $\ep(t)=0$. Important examples of such objectives
include
\begin{itemize}
 \item Transition probability
\[J_{\rmi\to\rmf}(\ep)=|\langle\rmf| U^\ep_T|\rmi\rangle|^2\]
This objective is maximized by a control which completely transfers the initial
state $|\rmi\rangle$ into the desired final state $|\rmf\rangle$.
 \item Expectation of a system observable $O$
\[J_O(\ep)=\Tr[U^{\ep\vphantom{\dagger}}_T\rho_0 U^{\ep\dagger}_T O]\]
 Here $O$ is a Hermitian matrix representing the observable and $\rho_0$ is
the initial system density matrix. The objective is maximized by
a
control which maximizes
quantum-mechanical average of $O$ at time $T$.
 \item Generation of a unitary process $W$
 \[
J_W(\ep)=\frac{1}{4}|\Tr(W^\dagger U^\ep_T)|^2\]
 Here $W$ is the unitary matrix representing a desired system evolution or a
desired quantum gate, for example Hadamard gate. Maximum of this objective is
achieved by a control such that $U_T=\rme^{i\phi} W$, where $\phi$ is arbitrary
(generally
unphysical) phase. Factor $1/4$ is chosen to have $\max\limits_\ep
J_W(\ep)=1$.
\end{itemize}

\noindent\textbf{Proof of the main result.} We will consider first
$J(\ep)=J_{\rmi\to\rmf}(\ep)$. For brevity, we will sometimes omit the
superscript $\ep$ in $U^\ep_t$ and $U^\ep_T$, and without loss of generality set
$\Delta=1$. Gradient
of $J(\ep)=|\langle\rmf| U^\ep_T|\rmi\rangle|^2$ for the
LZ system has the form~\cite{PechenTannor2011}
\begin{equation}\label{eq1:3}
\nabla J_\ep(t)=2\Im\Bigl(
\langle\rmi|U^\dagger_T|\rmf\rangle\langle\rmf|U_T^{\vphantom{\dagger}}
U^\dagger_t\sigma_zU^{\vphantom{\dagger}}_t|\rmi\rangle\Bigr)
\end{equation}
It can be written as $\nabla
J_\ep(t)=L(U^\dagger_t\sigma_zU^{\vphantom{\dagger}}_t)=l(t)$, where
$L:\su(2)\to\mathbb R$ is the
linear map on the Lie algebra of traceless Hermitian $2\times2$ matrices defined
by $L(A)=2\Im[
\langle\rmi|U^\dagger_T|\rmf\rangle\langle\rmf|U_T^{\vphantom{\dagger}}
A|\rmi\rangle]$, and $l(t)$ is a
real-valued function. If $\ep$ is a critical control field, then
$l(t)\equiv 0$
and therefore, in particular, $l'(t)=l''(t)=0$. These derivatives can be
computed to be
\begin{eqnarray*}
 l'(t) &=&L(-iU^\dagger_t[\sigma_x+\ep(t)\sigma_z,\sigma_z]
            U^{\vphantom{\dagger}}_t)
         =-2L(U^\dagger_t\sigma_y U^{\vphantom{\dagger}}_t)\\
 l''(t)&=&-2L(-iU^\dagger_t[\sigma_x+\ep(t)\sigma_z,\sigma_y]
 U^{\vphantom{\dagger}}_t)\\
       &&=-4L(U^\dagger_t\sigma_z U_t)+4\ep(t)L(U^\dagger_t\sigma_x
 U^{\vphantom{\dagger}}_t)
\end{eqnarray*}
Thus the condition $l''=l'=l=0$ for any $t$ such that $\ep(t)\ne 0$ takes
the form
\begin{equation}\label{eq2:2}
 L(U^\dagger_t\sigma_x U^{\vphantom{\dagger}}_t)= L(U^\dagger_t\sigma_y
 U^{\vphantom{\dagger}}_t)= L(U^\dagger_t\sigma_z U^{\vphantom{\dagger}}_t)=0
\end{equation}
The matrices $U^\dagger_t\sigma_x U^{\vphantom{\dagger}}_t$,
$U^\dagger_t\sigma_y U^{\vphantom{\dagger}}_t$, $U^\dagger_t\sigma_z
U^{\vphantom{\dagger}}_t$ are linearly independent traceless Hermitian $2\times
2$ matrices. They form a basis of $\su(2)$ and hence~(\ref{eq2:2}) implies
$L(A)=0$ for any $A\in\su(2)$.

Let $|\rmi_\bot\rangle$ be the state which is orthogonal to $|\rmi\rangle$.
Taking $A=|\rmi\rangle\langle \rmi_\bot|+ |\rmi_\bot\rangle\langle\rmi|$ and
$A'=i(|\rmi\rangle\langle\rmi_\bot|-|\rmi_\bot\rangle\langle\rmi|)$ gives
\begin{eqnarray*}
L(A)\hphantom{'}& =0\Rightarrow\Im \Bigl(\langle\rmi| U^\dagger_T
|\rmf\rangle\langle\rmf|U^{\vphantom{\dagger}}_T| \rmi_\bot\rangle\Bigr)=0\\
L(A')& =0\Rightarrow\Re \Bigl(\langle\rmi| U^\dagger_T
   |\rmf\rangle\langle\rmf|U^{\vphantom{\dagger}}_T|\rmi_\bot\rangle\Bigr)=0
\end{eqnarray*}
Thus $\langle\rmi|U^\dagger_T |\rmf\rangle\langle\rmf|U_T|\rmi_\bot\rangle=0$,
i.e. either $\langle\rmi|U^\dagger_T |\rmf\rangle=0$ or $\langle
\rmf|U_T|\rmi_\bot\rangle=0$. The former case corresponds to the global minimum
of the objective ($J=0$) and the latter to its global maximum ($J=1$). These are
the only allowed critical controls except of $\ep(t)\equiv0$. This finishes the
proof of the main result for $J_{\rmi\to\rmf}(\ep)$.

The analysis above immediately implies that if a linear map
$L:\su(2)\to\mathbb R$ satisfies
$L(U^\dagger_t\sigma_zU^{\vphantom{\dagger}}_t)=0$ then $L\equiv
0$. Now we will show that it means that the map $\chi:\ep\to U^\ep_T$ is
non-degenerate everywhere outside of $\ep(t)\equiv 0$. Since we
consider objectives produced by functions on $SU(2)$ which therefore
invariant with respect to the overall phase of $U^\ep_T$, we can
identify $U^\ep_T$ with the corresponding element of $SU(2)$. Small
variations
around $U^\ep_T$ can be represented as $\tilde U^\ep_T=U^\ep_T\rme^{\delta
w}\approx U^\ep_T(1+\delta w)$, where $\delta w=-\rmi\int_0^T
U^\dagger_t\sigma_zU^{\vphantom{dagger}}_t\delta\ep(t)dt$. For the map $\chi$ to
be non-degenerate,
$\tilde
U^\ep_T$ should span a neighborhood of $U^\ep_T$ that in turn requires $\delta
w$ to span $\su(2)$. If $\delta w$ does not span $\su(2)$, then there exists
$A\in\su(2), A\ne 0$ such that $(A,\delta w)\equiv\Tr (A^\dagger\delta w)=0$ for
all
$\delta\ep$ and hence
$L_A(U^\dagger_t\sigma_zU^{\vphantom{\dagger}}_t):=\Tr(A^\dagger
U^\dagger_t\sigma_zU^{\vphantom{\dagger}}_t)=0$. This is possible only if $A=0$
and hence the map
can not be degenerate. Therefore our result immediately
implies the absence of traps at any $\ep\ne 0$ for any objective functional
$J(U_T^\ep)$ which has no traps if considered as a function on $SU(2)$.
This includes
important objectives $J_O=\Tr[U^{\ep\vphantom{\dagger}}_T\rho_0 U^{\ep\dagger}_T
O]$ for maximizing
expectation of a system observable $O$ and $J_W=(1/4)|\Tr(W^\dagger U^\ep_T)|^2$
for
optimal generation of a unitary process $W$ (e.g., for unitary gate
generation). These objectives appear to be trap-free for the LZ system since
functions $f_O(U)=\Tr[U\rho_0U^\dagger O]$ and
$f_W(U)=(1/4)|\Tr(W^\dagger U)|^2$ have no local maxima on
$SU(2)$~\cite{NOTRAPS1}.

The control $\ep(t)\equiv0$ requires a separate consideration since the
condition $l''(t)=0$ for $\ep\equiv0$ can not be used to conclude
$L(U^\dagger_t\sigma_x U^{\vphantom{\dagger}}_t)=0$. This control is however not
a trap for example for $J_{\rmi\to\rmf}$ as shown by direct computation in the
Appendix.

\section{Discussion}
Now we discuss important
limitations for the present analysis. No
real-world system will perfectly evolve according to Eq.~(\ref{eq1:2}) and three
general kinds of deviations from the ideal situation include decoherence
effects, deviations of the actual control from the intended one
due to noise or imperfections of the laboratory setup,
and limited tunability of the control strength and time scales in
laboratory experiments. While we consider the system as evolving according to
the Schr\"odinger equation with unitary evolution, in real circumstances it can
experience additional influence of the environment which causes the dynamics to
be non-unitary. We also assume that any shape of the control $\ep(t)$ is
available, whereas typical pulses are either piecewise constant
or finite sums of cosines and sines at certain fixed frequencies.
These assumptions are common for the first step of control
landscape analysis which deals with the ideal
situation of noiseless unconstrained controls.
The next step upon establishment of the ideal landscape properties is to study
the effects of possible deviations, which we discuss below for the LZ system.

\begin{figure}[t]
\includegraphics[width=8.3cm]{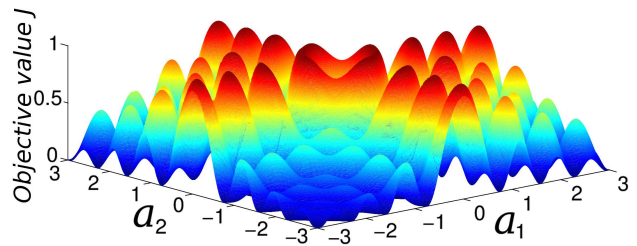}\caption{\label{fig2} (Color online) The
control landscape of $J_{0\to 1}(a_1,a_2)$ for the LZ system controlled by
piecewise constant controls ($N=2$, $T=10$, $\Delta=1$). The landscape
possesses multiple traps (local maxima).}
\end{figure}

The requirements on the available control fields (e.g. on their strength and
time scale) necessary for the conclusion of the absence of traps for attaining
maximal objective value are such that the available controls are sufficient to
guarantee controllability of the system. Minimal control time for the LZ system
can be estimated using the fundamental theory of optimal control at the quantum
speed limit as $T_{\rm QSL}\approx \Delta
E_0^{-1}\arccos(|\langle\rmi,\rmf\rangle|)$, where $\Delta E_0$ is the energy
variance of the free Hamiltonian $H_0=\Delta\cdot\sigma_x$ calculated on the
initial state~\cite{Caneva2009}. Hence our analysis applies to any final time
$T\gtrsim\pi\Delta E_0^{-1}$. If for a given physical system decoherence effects
occur on a time scale slower than $\Delta E_0^{-1}$, they can be neglected when
the control is implemented in the time optimal fashion. This shows that while
finite-time~\cite{Vitanov1996} and
decoherence~\cite{Ao1989,Akulin1992,Scala2011} effects can be important for the
LZ system, they do not modify the trap-free landscape property as soon as final
time $T$ is sufficiently smaller that the relaxation time and at the same time
is not too small to violate controllability of the systems.

\begin{figure}[t]
\includegraphics[scale=.4]{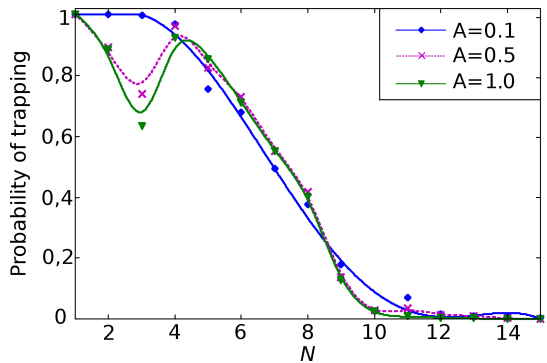}\caption{\label{fig3} (Color
online) Probability of trapping as a function of $N$ for piecewise constant
controls ($T=10, \Delta=1$). For every point, $10^3$ runs of MATLAB realization
of the Broyden-Fletcher-Goldfarb-Shanno (BFGS) optimization algorithm where
performed each starting at a random initial control $a=(a_1,\dots,
a_N)$~\cite{MATLAB}. Initial control amplitudes are uniformly distributed in the
range $a_i\in[A,A]$ but are allowed to escape this range during the search. The
search is defined as trapped if the attained objective is less than $0.99$.
Trapping may occur due to the presence of local maxima and/or principal
impossibility of attaining the objective value greater that $0.99$ with
available controls. Probability of trapping is estimated as a
fraction of trapped runs among all $10^3$ runs.}
\end{figure}

An extensive numerical analysis of control landscapes for multi-level model
systems with realistic laboratory control fields is provided
in~\cite{NumericalResults}. To analyze the role of limitations on the available
control fields for the LZ system, we numerically estimate the probability of
trapping when available controls are piecewise constant controls of the form
$\ep(t)=\sum_{i=1}^{N} a_i\chi_{[t_i,t_{i+1}]}(t)$, where
$\chi_{[t_i,t_{i+1}]}(t)=1$ if
$t\in[t_i,t_{i+1}]$ and zero otherwise and $a_i$ are the control parameters.
Typically,
$N\approx 100$ and control amplitudes are constrained within certain ranges, say
$a_i\in[-A, A]$. Exact solution for piecewise constant controls can be obtained
for example in the simplest case $N=1$. The objective for maximizing the
probability of
spin flip $J_{0\to 1}=|\langle 0|U^\ep_T|1\rangle|^2$ by a constant control
$\ep(t)=a$ can be computed to be $J_{0\to 1}(a)=\sin^2(T\sqrt{1+a^2})/(1+a^2)$.
Its traps (local maxima) are given by solutions of the equation
$\tan(T\sqrt{1+a^2})=T\sqrt{1+a^2}$; the corresponding objective values are
$J_{0\to 1}(a)=T^2/(1+T^2+T^2a)$. Control landscapes for a two-dimensional
control space ($N=2$) are more complex. Fig.~\ref{fig2} shows as an example the
control landscape of $J_{0\to 1}(a_1,a_2)$. The landscape has
multiple local maxima showing that significant restrictions on the control space
in an originally trap-free system may produce traps. Fig.~\ref{fig3} provides
the numerically estimated probability of trapping for piecewise constant
controls as a function of $N$. The probability of trapping becomes negligible
already for $N=10$--$15$ that means that limitations on the number of components
$N$
of available laboratory control fields have minor effect already for $N\gtrsim
10$ and hence should be negligible for realistic case $N\approx 100$.

In the laboratory, actual controls may deviate from the designed
numerically optimal pulse due to
noise and imperfections of experimental
setup~\cite{Noise}. These noise effects can influence on the landscape
structure by decreasing the maximal objective value.
We adopt
the general theory of~\cite{Moore2011} to analyze this influence for
the LZ system. Let $\ep_0(t)$ be an optimal control in the ideal situation of
absence of noise. In the presence of a random noise $\xi(t)$, the actual control
will fluctuate as $\ep(t)=\ep_0(t)+\varrho(t)\xi(t)$, where $\varrho(t)=1$ for
additive noise and $\varrho(t)=\ep_0(t)$ for multiplicative noise. A weak noise
modifies the averaged objective as
\[
\mathbb E[J(\ep_0)]\approx J(\ep_0)+\frac{1}{2}\int\limits_0^T\int\limits_0^T
{\cal H}_0(t,t')\varrho(t)\varrho(t')\mathbb E[\xi(t)\xi(t')] dtdt'
\]
where ${\cal H}_0(t,t')=\frac{\delta^2 J}{\delta\ep_0(t)\delta\ep_0(t')}$ is the
Hessian of the objective computed at the optimal control field $\ep_0$ and
$\mathbb E[\xi(t)\xi(t')]$ is the autocorrelation function of the noise. Since
Hessian is negative semidefinite at the maximum, the noise generally decreases
the average fidelity. The objective for additive (AWN) and multiplicative (MWN)
white noise with autocorrelation function $\mathbb
E[\xi(t)\xi(t')]=\sigma\delta(t-t')$, where $\sigma^2$ is the variance of the
noise amplitude distribution, takes the forms
\begin{eqnarray*}
\mathbb E_{\rm AWN}[J(\ep_0)]&\approx&
J(\ep_0)+\frac{\sigma^2}{2}\int\limits_0^T
{\cal H}_0(t,t)dt\\
\mathbb E_{\rm MWN}[J(\ep_0)]&\approx&
J(\ep_0)+\frac{\sigma^2}{2}\int\limits_0^T
{\cal H}_0(t,t)|\ep_0(t)|^2dt
\end{eqnarray*}
The last term in these equations is the noise-induced decrease $-{\cal
D}(\ep_0,\sigma,T)$ of the objective (such that $\mathbb E(J)\approx
J(\ep_0)-{\cal D}$ with ${\cal D}\ge 0$). The diagonal of the Hessian for
$J=J_{\rmi\to\rmf}$ can be shown to be ${\cal
H}_0(t,t)=-2|\langle\rmi|U^\dagger_t\sigma_z
U^{\vphantom{\dagger}}_t|\rmi_\bot\rangle|^2$ so that $|{\cal H}_0(t,t)|\le2$.
Therefore ${\cal D}(\ep_0,\sigma,T)$ for $J=J_{\rmi\to\rmf}$ is majorized by
\begin{eqnarray*}
{\cal D}_{\rm AWN}(\ep_0,\sigma,T)&\le&\sigma^2T\\
{\cal D}_{\rm MWN}(\ep_0,\sigma,T)&\le&\sigma^2 E
\end{eqnarray*}
where $E=\int_0^T|\ep_0(t)|^2dt$ is the total energy of the pulse. The diagonal
of the Hessian for the objective $J_W$ is ${\cal H}_0(t,t)=-2$ and therefore for
this objective ${\cal D}_{\rm AWN}=\sigma^2 T$ and ${\cal D}_{\rm
MWN}=\sigma^2 E$. It then follows that in both cases the influence of a weak
AWN can be minimized by using time optimal controls, while minimizing weak MWN
can be done by selecting less energetic pulses among all optimal pulses.

If the ideal landscape has multiple global optima with different
${\cal H}_0(t,t)$, then the noise induced decrease of objective can be
different at different optima that can produce traps in the non-ideal landscape.
Weak decoherence operates similarly to weak noise and can also produce
traps in the ideally trap-free landscape~\cite{NOTRAPS1}. These deviations from
the ideal situation should be avoided to reveal the trap-free landscape property
by either operating in the time optimal regime or using weak optimal controls
to combat MWN. Strong noise and strong decoherence that can significantly
modify the landscape are out of scope of this discussion.

{\it Conclusions.}---This work shows that unconstrained manipulation in the
Landau-Zener system is free of traps and hence unconstrained local search for
optimal controls is always able to find best optima. The impact on this result
of laboratory limitations due to decoherence, noise in the actual control
pulses, and restrictions on the available control fields is discussed.

\begin{acknowledgments}
A. Pechen acknowledges support of the Marie Curie International Incoming
Fellowship within the 7th European Community Framework Programme. N. Il'in
is partially supported by the Russian Foundation for Basic Research. This
research is made possible in part by the historic generosity of the Harold
Perlman family and by the Ministry of Education and Science of the Russian
Federation, project 8215.
\end{acknowledgments}

\section*{Appendix}
Here we prove that the control $\ep(t)\equiv0$ is not a trap for
state-to-state transfer described by the
objective $J(\ep)=J_{\rmi\to\rmf}(\ep)$. The
evolution
operator
produced by $\ep(t)=0$ has the form $U_t=\rme^{-it\sigma_x}$. Therefore
$V_t:=U^\dagger_t\sigma_z U^{\vphantom{\dagger}}_t=\cos(2t)\sigma_z+\sin
(2t)\sigma_y$ and the gradient of the objective is
\[
\nabla J_{\ep=0}(t)=\cos 2t\cdot L(\sigma_z)+\sin 2t\cdot L(\sigma_y)
\]
If $\ep(t)=0$ is a critical point, then $\nabla J_{\ep=0}(t)=0$ for any
$t\in[0,T]$, and hence $L(\sigma_z)=L(\sigma_y)=0$. If $\alpha:=L(\sigma_x)=0$,
then $L\equiv 0$ on $\su(2)$ and similarly to the proof of the main result we
conclude that $\ep=0$ is not a trap.

Now consider the case $\alpha\ne 0$. In this case $|\rmi\rangle$ and
$|\rmf\rangle$ are such that $\ep=0$ is neither a global maximum nor global
minimum. The evolution operator produced by a small variation of the control
$\dl\ep$ can be represented as $U^{\dl\ep}_T=\rme^{-i T\sigma_x}W_T$, where
$W_T$ satisfies
\[
 \dot W_t=-\rmi\dl\ep(t)V_tW_t,\qquad W_0=\mathbb I
\]
The operator $W_T$ can be computed up to the second order in $\dl\ep$ as
\begin{eqnarray*}
 W_T&=&\mathbb I+A_1+A_2+o(\|\dl\ep\|^2)\\
 && A_1=-\rmi\int_0^T dt\dl\ep(t)V_t,\\
 && A_2=-\int_0^Tdt_1\int_0^{t_1}dt_2\dl\ep(t_1)\dl\ep(t_2)
V_{t_1}V_{t_2}
\end{eqnarray*}
This gives the perturbation expansion for the objective
(here $|\rmf'\rangle=\rme^{\rmi T\sigma_z}|\rmf\rangle$)
\begin{eqnarray*}
J(\dl\ep)&=& |\langle\rmf'|\mathbb I+A_1+A_2+\dots|\rmi\rangle|^2\\
&=&
|\langle\rmf'|\rmi\rangle|^2+\dl J_1(\dl\ep)+\dl J_2(\dl\ep)+o(\|\dl\ep\|^2)
\end{eqnarray*}
where
\begin{eqnarray*}
\dl J_1(\dl\ep)&=&
2\Re(\overline{\langle\rmf'|\rmi\rangle}\langle\rmf'|A_1|\rmi\rangle)\\
\dl J_2(\dl\ep)&=&|\langle\rmf'|A_1|\rmi\rangle|^2+
2\Re(\overline{\langle\rmf'|\rmi\rangle}\langle\rmf'|A_2|\rmi\rangle)
\end{eqnarray*}
Hence the variation of the objective satisfies (note that $|\langle
\rmf'|\rmi\rangle|^2=J(0)$)
\[
\dl J=J(\dl\ep)-J(0)=\dl J_1(\dl\ep)+\dl J_2(\dl\ep)+o(\|\dl\ep\|^2)
\]

If $\ep=0$ is a critical control, then $\dl J_1(\dl\ep)=0$ for any $\dl\ep$. We
will show the existence of controls $\dl\ep_1$ and
$\dl\ep_2$ such that $\dl
J_2(\dl\ep_1)$ and $\dl J_2(\dl\ep_2)$ have opposite signs. It is sufficient
to choose $\dl\ep_1$ and
$\dl\ep_2$ to satisfy $\langle \rmf'|A_1|\rmi\rangle=0$, e.g.
\[
 \int_0^T dt\dl\ep_i(t)\cos 2t=\int_0^T dt\dl\ep_i(t)\sin 2t=0,\qquad i=1,2.
\]
Since
$V_{t_1}V_{t_2}=\cos 2(t_1-t_2)+\rmi\sigma_x\sin 2(t_1-t_2)$, we have
\begin{eqnarray*}
\dl J_2(\dl\ep)&=& 2\int_0^Tdt_1\int_0^{t_1}
dt_2\dl\ep(t_1)\dl\ep(t_2)\Bigl(J(0)\cos2(t_1-t_2)\\
&& +\alpha\sin2(t_1-t_2)\Bigr)\\
&=& 2\alpha\int_0^Tdt_1\int_0^{t_1}dt_2\dl\ep(t_1)\dl\ep(t_2)\sin2(t_1-t_2)
\end{eqnarray*}
Assuming for simplicity that $T\ge \pi$, we take
$\dl\ep_1(t)=\chi_{[0,\pi]}(t)$ and $\dl\ep_2(t)=\cos
(4t)\chi_{[0,\pi]}(t)$, where $\chi_{[0,\pi]}(t)$ is the characteristic
function of the interval $[0,\pi]$. Then $\dl J_2(\dl\ep_1)=-\pi\alpha$ and
$\dl J_2(\dl\ep_2)=\pi\alpha/6$. Therefore for $\alpha\ne 0$ there exist
control variations around $\ep(t)=0$ increasing the objective and
control variations decreasing it. This implies that $\ep(t)=0$ is neither a
local maximum nor minimum.


\begin{thebibliography}{99}
\bibitem{MATLAB} We use BFGS Hessian update for choosing the search
direction in the
quasi-Newton algorithm built in the MATLAB routine fminunc for unconstrained
nonlinear optimization. Exact gradient of the objective is provided.
Default MATLAB stopping criteria are modified as follows: maximum number of
objective evaluations is $10^3 N$, termination tolerance on the
objective value and on $a_i$ is $10^{-8}$.
\end{thebibliography}
\end{document}